\documentclass{article}
\usepackage{graphicx}
\usepackage{emulateapj,timesfonts}

\def\emph#1{{\em #1}}
\def\***#1{{\sc *** #1 ***}}

\let\ROSAT\rosat
\def\LOG{\mathop{\rm LOG}\nolimits}
\def\ergpersec{ergs~s$^{-1}$}
\def\ergs {ergs$\;$s$^{-1}\,$cm$^{-2}$}

\submitted{ApJ Letters, in press}

\begin{document}
\lefthead{CLUSTER EVOLUTION}
\righthead{VIKHLININ ET AL.}

\title{EVOLUTION OF CLUSTER X-RAY LUMINOSITIES AND RADII: 
	RESULTS FROM THE 160 SQUARE DEGREE \ROSAT\/ SURVEY}
\author{A.\ Vikhlinin\altaffilmark{1}\altaffilmark{,5}, 
B.\ R.\ McNamara\altaffilmark{1}, W.\ Forman\altaffilmark{1},
C.\ Jones\altaffilmark{1}, H.\ Quintana\altaffilmark{2,3}, and
A.\ Hornstrup\altaffilmark{4}}

\altaffiltext{1}{Harvard-Smithsonian Center for Astrophysics, 60 Garden St.,
Cambridge, MA 02138; avikhlinin@cfa.harvard.edu}
\altaffiltext{2}{Dpto.\ de Astronomia y Astrofisica, Pontificia Universidad
Catolica, Casilla 104, 22 Santiago, Chile}
\altaffiltext{3}{Presidential Chair in Science}
\altaffiltext{4}{Danish Space Research Institute, Juliane Maries Vej 30,
2100  Copenhagen O, Denmark}
\altaffiltext{5}{Also Space Research Institute, Moscow, Russia}

\begin{abstract}
We searched for cluster X-ray luminosity and radius evolution using our
sample of 201 galaxy clusters detected in the 160~deg$^2$ survey with the
\ROSAT\/ PSPC (Vikhlinin et al.\ 1998). With such a large area survey, it is
possible, for the first time with \ROSAT\/, to test the evolution of
luminous clusters, $L_x>3\times10^{44}\,$\ergpersec\ in the 0.5--2~keV band.
We detect a factor of 3--4 deficit of such luminous clusters at $z>0.3$
compared to the present.  The evolution is much weaker or absent at modestly
lower luminosities, 1--$3\times10^{44}\,$\ergpersec. At still lower
luminosities, we find no evolution from the analysis of the $\log N -
\log S$ relation. The results in the two upper $L_x$ bins are in agreement with
the {\em Einstein\/} EMSS evolution result (Gioia et al.\ 1990a, Henry et
al.\ 1992) while being obtained using a completely independent cluster
sample.  The low-$L_x$ results are in agreement with other \ROSAT\/ surveys
(e.g.\ Rosati et al.\ 1998, Jones et al.\ 1998).

We also compare the distribution of core radii of nearby and distant
($z>0.4$) luminous (with equivalent temperatures 4--7~keV) clusters, and
detect no evolution. The ratio of average core radius for $z\sim0.5$ and
$z<0.1$ clusters is $0.9\pm0.1$, and the core radius distributions are
remarkably similar. A decrease of cluster sizes incompatible with our data
is predicted by self-similar evolution models for high-$\Omega$ universe.

\end{abstract}

\keywords{galaxies: clusters: general --- surveys --- X-rays: galaxies}

\section{Introduction}

The cluster evolution rate is a strong test of cosmological parameters
(e.g., White \& Rees 1978, Kaiser 1986, Eke, Cole \& Frenk 1996). It is best
to study evolution using X-ray selected samples of distant clusters which
are much less affected by projection than the optically selected samples
(van Haarlem et al.\ 1997).  Of all the interesting cluster parameters such
as mass, velocity dispersion, and temperature, the X-ray luminosity is the
most accessible to measurements with present-day instruments, and most of the
earlier studies were focused on evolution of the cluster X-ray luminosity
function.

A strong evolution of cluster luminosities at $z\sim0.1$ was reported from
the EXOSAT survey (Edge et al.\ 1990), but was later disproved by the
\ROSAT\/ All-Sky Survey (Ebeling et al.\ 1997). At higher redshifts,
negative evolution of the cluster X-ray luminosity function was first
reported by Gioia et al.\ (1990a) using the {\em Einstein}\/ Extended Medium
Sensitivity Survey (EMSS; Gioia et al.\ 1990b, Stocke et al.\ 1991). Gioia
et al.\ and later Henry et al.\ (1992) compared the cluster luminosity
functions below and above $z=0.3$.  They found that while the number of the
low luminosity clusters does not evolve, there is a significant deficit of
luminous, $L_x(0.3-3.5\mbox{~keV})>5\times10^{44}\,$\ergpersec, clusters at
high redshift.

This EMSS result was questioned recently. Nichol et al.\ (1997) reanalyzed
the EMSS cluster sample using \ROSAT\/ X-ray and new optical observations
and argued that the evolution reported in the original EMSS papers was not
significant.  Several groups pursued independent searches for distant
clusters in archival \ROSAT\/ PSPC observations. Collins et al.\ (1997)
found that the redshift distribution of 35 clusters detected in their
17~deg$^2$ survey is consistent with no evolution.  This contradicted the
earlier claim by Castander et al.\ (1995) of a strong evolution in a similar
sample; however, the latter authors used an X-ray source detection algorithm
not optimized for the cluster search.  Jones et al.\ (1998) presented the
$\log N - \log S$ relation for 46 clusters from their 16~deg$^2$ survey and
found that this relation is consistent with no evolution of the
$L_x<2\times10^{44}\,$\ergpersec\ (0.5--2~keV band) clusters.  Rosati et
al.\ (1998) derived cluster luminosity functions up to $z\sim 0.8$ from
their sample of 70 clusters detected in a 33~deg$^2$ survey, and found no
evolution at low luminosities, $L_x<3\times10^{44}\,$\ergpersec. However,
none of these \ROSAT\/ surveys covers an area large enough to probe the
evolution of the luminous clusters, and their no-evolution claims do not
contradict the EMSS results.

Our 160~deg$^2$ survey (Vikhlinin et al.\ 1998, hereafter Paper~I) is the
first \ROSAT\/ survey comparable with the EMSS in sky coverage for distant
clusters. We are able to test, and confirm, the EMSS evolution results even
with the incomplete redshift data currently at hand. Eventually, when the
spectroscopic work is complete, we will be able to characterize the
luminosity evolution more accurately. In this Letter, we also show that the
cluster X-ray core radii do not evolve between $z\sim0.5$ and now. Throughout
the paper, we use definitions $f_{-14}$ and $L_{44}$ for flux and luminosity
in the 0.5--2~keV energy band in units of $10^{-14}\,$\ergs\ and
$10^{44}\,$\ergpersec, respectively. We also use
$H_0=50$~km~s$^{-1}$~Mpc$^{-1}$ and $q_0=0.5$.

\section{Cluster Sample}

In Paper~I, we presented a catalog of 223 extended X-ray sources detected in
646 high Galactic latitude \ROSAT\/ PSPC observations. For each detected
source, we measured the X-ray flux and angular core-radius.  We optically
confirmed 89\% of detected sources as clusters of galaxies; 8\% are false
detections due to point source confusion and 3\% still lack optical data. In
the high X-ray flux range, which is the focus of the present work, 80 out of
82 detected sources are optically confirmed clusters. We measured and
compiled from the literature spectroscopic redshifts for 76 clusters.  For
the rest of the optically confirmed clusters, redshifts are estimated with
an accuracy of $\Delta z \approx 0.07$ by optical photometry of the
brightest cluster galaxies.  All the X-ray and optical data are presented in
Paper~I. Below we use these data to constrain the evolution of cluster
luminosities and sizes at high redshift, $z\sim 0.5$.

\medskip
\tabcaption{\centerline{All bright clusters lacking spectroscopic redshifts}
		\label{tab:listbright}}
{\footnotesize
\def\arraystretch{1.05}
\begin{center}
\begin{tabular}{crlcl}
\hline
\hline
Name & \multicolumn{1}{r}{$f_{-14}$$^{\rm a}$\phantom{l}} &
\multicolumn{1}{c}{$z$$^{\rm b}$} & $z$ range$^{\rm b}$ & 
\multicolumn{1}{c}{$z_{\rm min}$$^{\rm c}$} \\
\hline
2137+0026 &     27.8 & 0.05 & 0.00--0.12 &0.5\\
1301+1059 &     28.1 & 0.30 & 0.23--0.34 &0.5\\
0841+6422 &     29.1 & 0.36 & 0.29--0.40 &0.5\\
1722+4105 &     29.4 & 0.33 & 0.26--0.37 &0.5\\
1641+4001 &     29.4 & 0.51 & 0.44--0.55 &0.5 $+$\\
1524+0957 &     30.4 & 0.11 & 0.04--0.15 &0.5\\
0259+0013 &     32.4 & 0.17 & 0.10--0.21 &0.5\\
2348--3117&     32.5 & 0.21 & 0.14--0.28 &0.5\\
0159+0030 &     32.7 & 0.26 & 0.19--0.30 &0.5\\
1515+4346 &     34.6 & 0.26 & 0.19--0.30 &0.5\\
0050--0929&     36.6 & 0.21 & 0.14--0.25 &0.5\\
2319+1226 &     38.2 & 0.25 & 0.18--0.29 &0.5\\
1146+2854 &     39.2 & 0.17 & 0.10--0.21 &0.5\\
1124+4155 &     40.1 & 0.18 & 0.11--0.22 &0.5\\
0532--4614&     41.1 & 0.10 & 0.03--0.17 &0.5\\
1013+4933 &     45.6 & 0.17 & 0.10--0.21 &0.4\\
1142+2144 &     45.9 & 0.18 & 0.11--0.22 &0.4\\
0958+5516 &     48.2 & 0.20 & 0.13--0.24 &0.4\\
0237--5224&     64.4 & 0.13 & 0.06--0.20 &0.4\\
1418+2510 &     75.6 & 0.24 & 0.17--0.28 &0.4\\
1641+8232 &     80.5 & 0.26 & 0.19--0.30 &0.3 $+$\\
1206--0744&    129.0 & 0.12 & 0.05--0.16 &0.3\\
1630+2434 &    179.4 & 0.09 & 0.02--0.13 &0.3\\
\hline
\end{tabular}
\end{center}

\noindent $^{\rm a}$ X-ray flux in the 0.5--2~keV band, $10^{-14}\,$\ergs.

\noindent $^{\rm b}$ Photometric redshift and its $95\%$ confidence range.

\noindent $^{\rm c}$ Minimum redshift required for inclusion in the
high-$L_x$, high-$z$ subsample (see text); this redshift is defined by the
observed flux. Clusters, whose actual redshifts can possibly exceed this
minimum value, are marked $+$.

}
\medskip

\begin{figure*}[htb]
\hfill\includegraphics[width=3.5in]{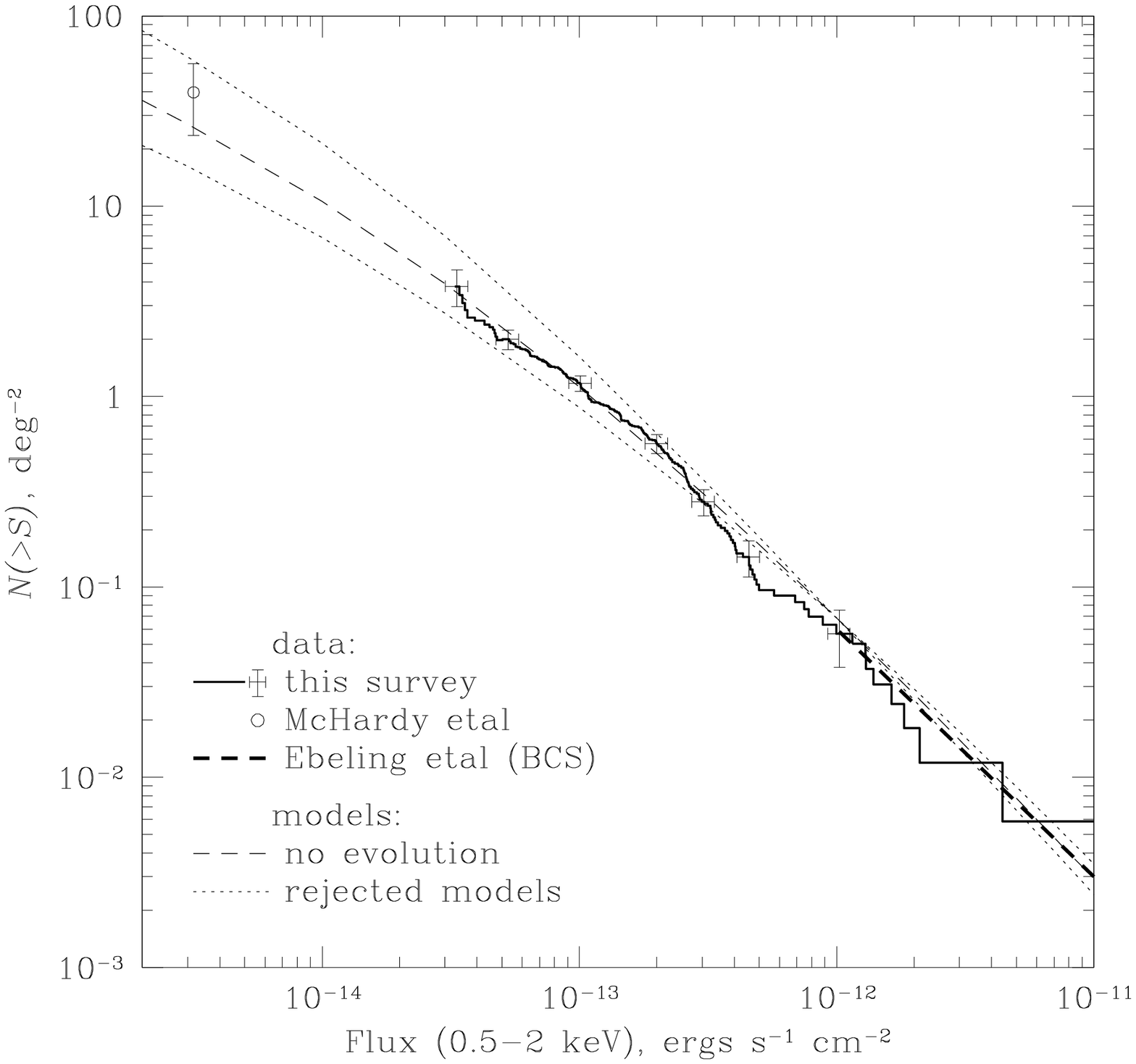}\hfill
      \includegraphics[width=3.5in]{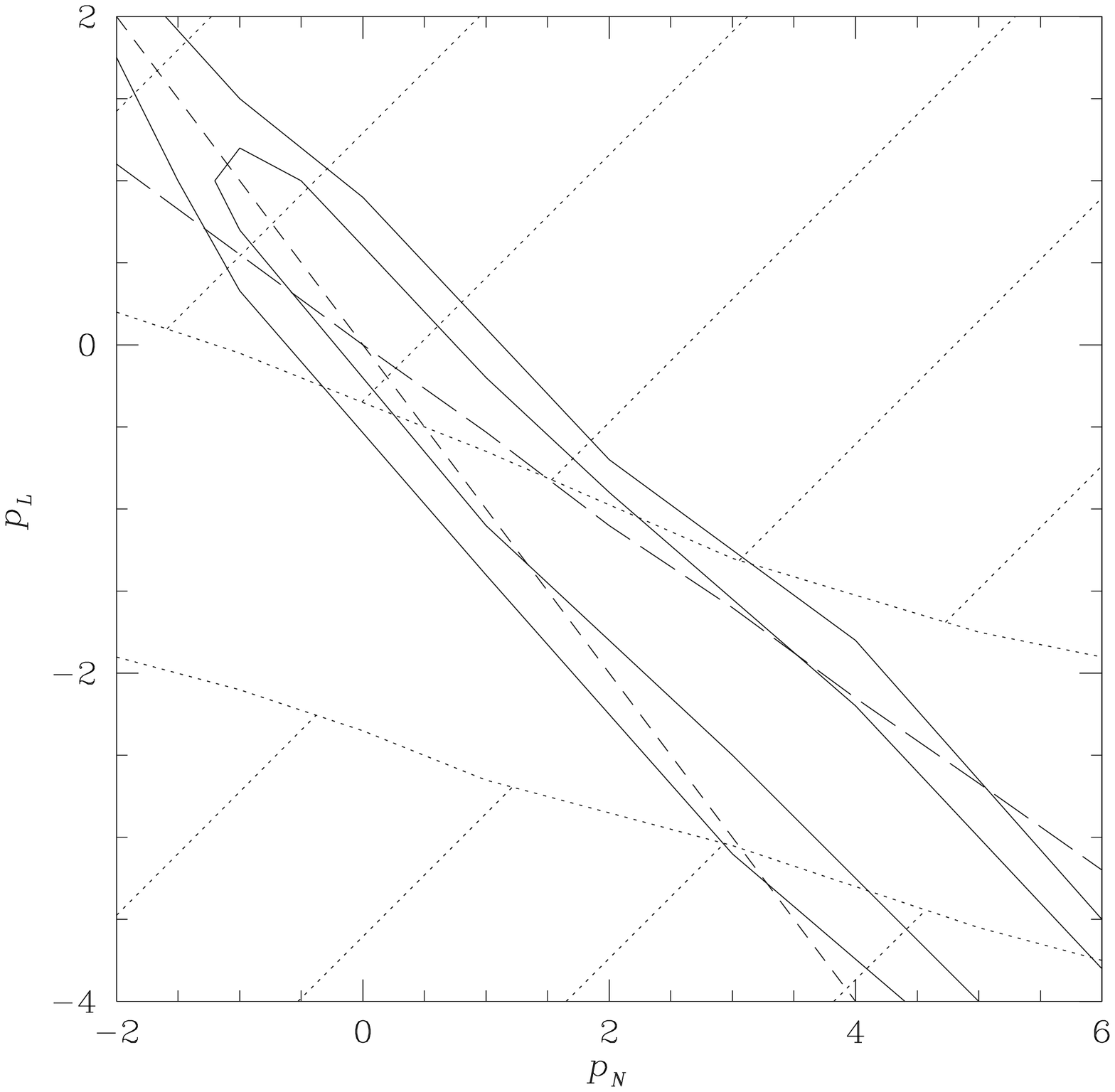}\hfill
\vskip -15pt
\caption{\footnotesize Fit to the $\log N - \log S$ relation. The observed
$\log N - \log S$ relation (\emph{a}) was fit by a family of evolution
models which combine luminosity and density evolutions as $(1+z)^{p_L}$ and
$(1+z)^{p_N}$, respectively. Solid lines in panel \emph{b}\/ show 68\% and
95\% confidence intervals of $p_N$ and $p_L$. The short- and long-dashed
lines in (\emph{b}) correspond to no-evolution of the comoving volume
emissivity of clusters at $z=0.5$, obtained by integration of the luminosity
function from 0 to $\infty$ (short dash) or from $10^{42}$ to
$10^{46}\,$\ergpersec\ (long dash).  Shaded regions in the right panel
correspond to parameters which predict either too many (top, $>6.3$) or too
few (bottom, $<0.36$) clusters in the high-$L_x$, high-$z$ sample. The
no-evolution model is shown as a dashed line in (\emph{a}). All allowed
models predict virtually the same $\log N - \log S$.  To illustrate the
accuracy of the fit, we show two rejected models, $p_N=0, p_L=-1$ and
$p_N=2,p_L=0$ (lower and upper dotted lines in the left panel,
respectively). Error bars show statistical uncertainties of the measured
$\log N - \log S$ at several representative fluxes. Note that the
no-evolution model derived by Rosati et al.\ (1998) differes from that shown
here by 10--20\% since they integrated only to $z=1.1$ and used a different
local luminosity function. }
\label{fig:lognlogs}
\end{figure*}

\section{Deficit of luminous clusters at high redshift}\label{sec:deficit}

Cluster evolution is detected in the EMSS sample only for the most luminous
clusters, $L_{44}>5$ in the 0.3--3.5~keV energy band, or $L_{44} \gtrsim 3$
in the 0.5--2 keV band (Gioia et al.\ 1990a, Henry et al.\ 1992). The
lower-luminosity clusters in the EMSS sample show little or no evolution. We
will search for evolution in our sample above this limiting luminosity.
Although we cannot derive accurate luminosity functions with the presently
incomplete spectroscopic data, a sample of luminous high-redshift clusters
can be selected using the observed X-ray flux and the conservative upper
bound of their estimated redshift.  With such a sample, one can test the
evolution of the cluster luminosity function by comparing the number of
detected clusters with the prediction of the no evolution model.

\subsection{High-Luminosity and High-Redshift Subsample}

A luminosity $L_{44}=3$ corresponds to observed fluxes, $f_{-14}=77.5$,
44.3, and 27.8 at redshifts $z=0.3$, 0.4, and 0.5, respectively.  Our
high-luminosity, high-redshift subsample is defined using these limiting
fluxes as follows. The cluster flux must be $f_{-14}>77.5$ and redshift must
be $z>0.3$, or $f_{-14}>44.3$ and $z>0.4$, or $f_{-14}>27.8$ and $z>0.5$.
For this sample definition, the lower limit of the luminosity varies with
redshift; the variations are, however, limited between $L_{44}=3$ and $6$
for $z<0.7$.

There are 48 clusters with fluxes $f_{-14}>27.8$ in our sample.
Spectroscopic redshifts are available for 25 of them; none of these 25
clusters satisfies the selection criteria above. The remaining 23 clusters
with photometric redshifts are listed in Table~\ref{tab:listbright}. Column
(4) in this Table shows the 95\% confidence interval of the photometric
redshift.  The observed flux corresponds to the minimum redshift required by
our sample definition (column~5). It can be seen that all clusters except
1641+4001 and 1641+8232 can be confidently excluded from the high-$L_x$,
high-$z$ subsample. We conclude that at most, only two clusters belong to
this subsmaple.

\subsection{Comparison with No-Evolution Model Predictions}

To calculate the expected number of observed clusters, we integrated the
local luminosity function (Ebeling et al.\ 1997) in the appropriate redshift
and luminosity range and accounted for the survey solid angle as a function
of flux (Paper I). We then compare these predictions with the observed
number of clusters in different subsamples. The results are presented in
Table~\ref{tab:noevol}. For the no-evolution model, we expect 9.3 clusters
in the high-$L_x$, high-$z$ subsample, where we observe at most 2.  Such a
deviation is significant at more than 99.5\% confidence. For the negative
evolution observed in the EMSS, we predict that this subsample should
contain 2--3 clusters, in agreement with the observed number.  Finally, for
$q_0=0$, the no evolution model predicts 8.1 clusters; the observed deficit
is still significant in this case.

For a consistency check, we compare the number of $z<0.3$ clusters above the
same limiting fluxes, $f_{-14}=77.5$, 44.3, and 27.8, with the prediction of
the no evolution model (Table~\ref{tab:noevol}); in all flux bins, there is
an excellent agreement.

\begin{center}
\tabcaption{\centerline{Comparison with no-evolution model
predictions\label{tab:noevol}}}
\footnotesize
\def\arraystretch{1.15}
\begin{tabular}{lccr}
\hline
\hline
\multicolumn{1}{c}{Subsample} & 
\multicolumn{1}{c}{Predicted} & 
\multicolumn{1}{c}{Observed}  & 
\multicolumn{1}{c}{Prob} \\
\hline
high-$L_x$, high-$z$           & 9.3     &  2 & $<0.005$ \\
$z<0.3$,~~ $f_{-14}>27.8$      & 32.5 & 39 & \nodata\\
$z<0.3$,~~ $f_{-14}>44.3$      & 20.4 & 21 & \nodata\\
$z<0.3$,~~ $f_{-14}>77.5$      & 11.3 & 10 & \nodata\\
$z>0.4$,~~ $f_{-14}>13.9$      & 22.2 & 18 & 0.22\\
\hline
\end{tabular}
\end{center}
\bigskip

We also can compare the observed and predicted number of lower-luminosity
clusters at high redshift. For that, we use a subsample of clusters with
fluxes in the range $13.9<f_{-14}<44.3$ and $z>0.4$. This flux range at
$z=0.4$ corresponds to the luminosity range $1<L_{44}<3$, which combines the
two lowest luminosity bins in the EMSS luminosity function at high redshift.
To obtain a conservative lower limit of the number of observed clusters, we
use the lower bound of photometric redshifts. We find 18 clusters, compared
with 22.2 predicted by the no-evolution model. That is, the evolution in
this interval is certainly different from that of the high $L_x$ clustrs.

To summarize, we find a large deficit, by a factor of 3--4, of $L_{44}>3$
clusters at high redshift, similarly to the EMSS result. The evolution rate
is smaller at lower luminosities, $1<L_{44}<3$, again similar to the EMSS
and other \ROSAT\/ surveys (Rosati et al.\ 1998). At still lower $L_x$, our
redshift database is very incomplete, but additional constraints on
cluster evolution can be derived from the $\log N - \log S$ relation,
as described below.

\section{$\LOG \;N\; -\; \LOG\; S\;$ relation for clusters}

The number of clusters as a function of flux (the $\log N - \log S$
relation) can provide some constraints on evolution without redshift
information.  To obtain the evolution constraints from the $\log N - \log
S$, we parameterize the evolution of the luminosity function as a
combination of pure luminosity and pure number density evolution, both as
powers of $(1+z)$.  With this type of evolution, the luminosity function at
redshift $z$ can be expressed through the local luminosity function $F_0(L)$
as:
\begin{equation}\label{eq:lfevol}
F_z(L) = (1+z)^{p_N}\, F_0\left(L/(1+z)^{p_L}\right).
\end{equation}
In this equation, $p_N$ and $p_L$ parameterize the rate of the density and
luminosity evolution, respectively. These parameters can be constrained by
fitting the observed $\log N - \log S$ relation.

To predict the model $\log N - \log S$ relation, we integrated the local
luminosity function (Ebeling et al.\ 1997), scaled according to
eq.~(\ref{eq:lfevol}) in the redshift interval $0<z<2$ and in the luminosity
interval $0.01<L_{44}<100$. The model $\log N - \log S$ was normalized by
the surface density of clusters expected in the no-evolution model above
$10^{-12}\,$\ergs, approximately the completeness limit of the Ebeling et
al.\ \ROSAT\/ All-Sky Survey sample. Finally, we multiplied the model $\log
N - \log S$ relation by the solid angle of our survey as a function of flux
(Paper I).

We then found the allowed range of parameters $p_N$ and $p_L$ using the
$C$-statistic (Cash 1979) calculated in the flux range ($f_{-14}$) from 100
to 4, where the solid angle of our survey still exceeds 4~deg$^2$.  The 68\%
and 95\% confidence region is shown by solid lines in the right panel of
Fig~\ref{fig:lognlogs}. The allowed combination of $p_N$ and $p_L$
corresponds approximately to a non-evolving comoving volume emissivity.  The
dotted and dashed lines in the right panel correspond to no evolution of the
volume emissivity of all clusters (i.e.\ defined by $p_N+p_L=0$), and
clusters in the range $10^{42}$--$10^{46}\,$\ergpersec\ at $z=0.5$,
respectively.  The no-evolution model ($p_N=0$, $p_L=0$) is shown by the
dashed line in the left panel of Fig~\ref{fig:lognlogs}. All the allowed
$p_N, p_L$ parameters predict virtually the same $\log N - \log S$
relations.  On the contrary, the rejected models predict $\log N - \log S$
relations which are markedly different from the data (dotted lines in the
left panel of Fig~\ref{fig:lognlogs}).

The observed $\log N - \log S$ relation favors no evolution of the cluster
volume emissivity. This indirectly implies that the luminosity function at
low $L_x$, which dominates the volume emissivity, does not evolve.  This,
however, does not imply that there is no evolution at all. For example, the
$p_N=4,p_L=-3$ model is allowed. Since the power-law slope of the luminosity
function is close to $-1.8$ (Ebeling et al.\ 1997), the number of low
luminosity clusters, $L_{44}<1$, does not evolve in this model, while the
number of $L_{44}>3$ clusters decreases significantly at $z>0.3$. This
behavior is similar to what we find in \S\ref{sec:deficit}.

\section{Evolution of cluster sizes}

In Paper I, we measured angular core radii by fitting surface brightness
distributions with the $\beta$-model (e.g., Jones \& Forman 1984). Here we
compare the radius distribution for our distant, $z>0.4$, clusters and for
nearby clusters in Jones \& Forman (1998) sample.  To convert angular radii
to proper sizes of distant, $z>0.3$, clusters, we used both spectroscopic
and photometric redshifts. The accuracy of the photometric redshift, $\pm
0.07$, is sufficient for this purpose. The corresponding proper size
uncertainty is 15\% at $z=0.3$, which is smaller than the statistical
uncertainty of the angular radius measurements, $\sim 20\%$. Jones \& Forman
fitted both core-radius and $\beta$ while we fixed $\beta=0.67$ for distant
clusters. For consistency, we converted core radii from Jones \& Forman to
$\beta=0.67$ using Eq.~(4) of Paper I. Since Jones \& Forman found a
correlation between core-radius and luminosity, we matched the luminosity
ranges by using only clusters with $1<L_{44}<5$ (corresponding to
temperatures 4--7~keV) in both samples.  The luminosities of distant
clusters were computed using both spectroscopic and photometric redshifts.
The accuracy of photometric redshifts is sufficient because radius is a weak
function of luminosity. The median redshift of the 25 selected distant
clusters is $z_{\rm med}=0.51$. The uncertainty of an individual radius
measurement is $\lesssim 30\%$, which includes photon statistics, background
modeling, and scatter in $\beta$-parameters (Paper I).  This is
significantly smaller than the intrinsic width of the derived radius
distributions.

The core radius distributions for distant and nearby clusters are remarkably
similar (Fig.~\ref{fig:axdistr}), especially for $q_0=0$. To characterize
the radius evolution quantitatively, we use the ratio of median radii in
both samples. The median radius is $240\pm14$~kpc for nearby clusters, and
$210\pm14$~kpc ($q_0=0.5$) or $230\pm16$~kpc ($q_0=0$) for distant clusters.
The ratio lies in the range $0.95-1.25$.  Some models predict a much
stronger evolution of radii.  For example, in Kaiser's (1986) self-similar
models, the cluster radius grows by a factor of $\sim 2$ from $z=0.4$ to the
present, while hydrodynamic simulations with $\Omega+\Lambda=1$ (Cen
\& Ostriker 1994) show a factor of 1.5 growth. However, neither these models
nor our measurements account for cooling flows (e.g.\ Fabian 1994) which can
cause underestimation of the core-radius. A comparison of core radii is not
meaningful if the cooling flow fraction changes with redshift.

\section{Conclusions}

We present a first \ROSAT\/ analysis of the evolution of luminous,
$L_x>3\times10^{44}\,$\ergpersec\ distant clusters. We find a significant,
factor of 3--4, decrease in the number of such clusters at $z>0.3$,
confirming the detection of evolution in the EMSS (Gioia et al.\ 1990a,
Henry et al.\ 1992). At lower luminosities, 1--$3\times10^{44}\,$\ergpersec,
the evolution is undetectable, with a decrease in number by a factor of only
$1.3\pm0.2$.  This is also consistent with the EMSS and other \ROSAT\/
surveys (e.g.\ Rosati et al.\ 1998). The absence of evolution of low
luminosity clusters is also supported by the analysis of the $\log N - \log
S$ distribution from which we find that the cluster volume emissivity,
dominated by low-luminosity objects, does not evolve. The observed evolution
can be reproduced by a model in which the characteristic luminosity
decreases with redshift, but the comoving number density of clusters
increases. Such models arise naturally in the hierarchical cluster formation
scenario (e.g.\ Kaiser 1986).

We compare the distribution of core-radii of distant, $z>0.4$ and nearby
clusters.  We find that the distribution of core radii at $z>0.4$ is very
similar to that in nearby clusters; the average radius has changed at
$z>0.4$ by a factor of only $0.9\pm0.1$. A stronger change is expected for
hierarchical cluster formation in a flat universe (Kaiser 1986, Cen \&
Ostriker 1994). We also note that the assumption of no evolution of cluster
sizes has been essentially used in flux measurements and area calculations
in several X-ray surveys (e.g.\ EMSS, Nichol et al. 1997), but is only
verified here for the first time.

\bigskip
\centerline{\includegraphics[width=3.25in]{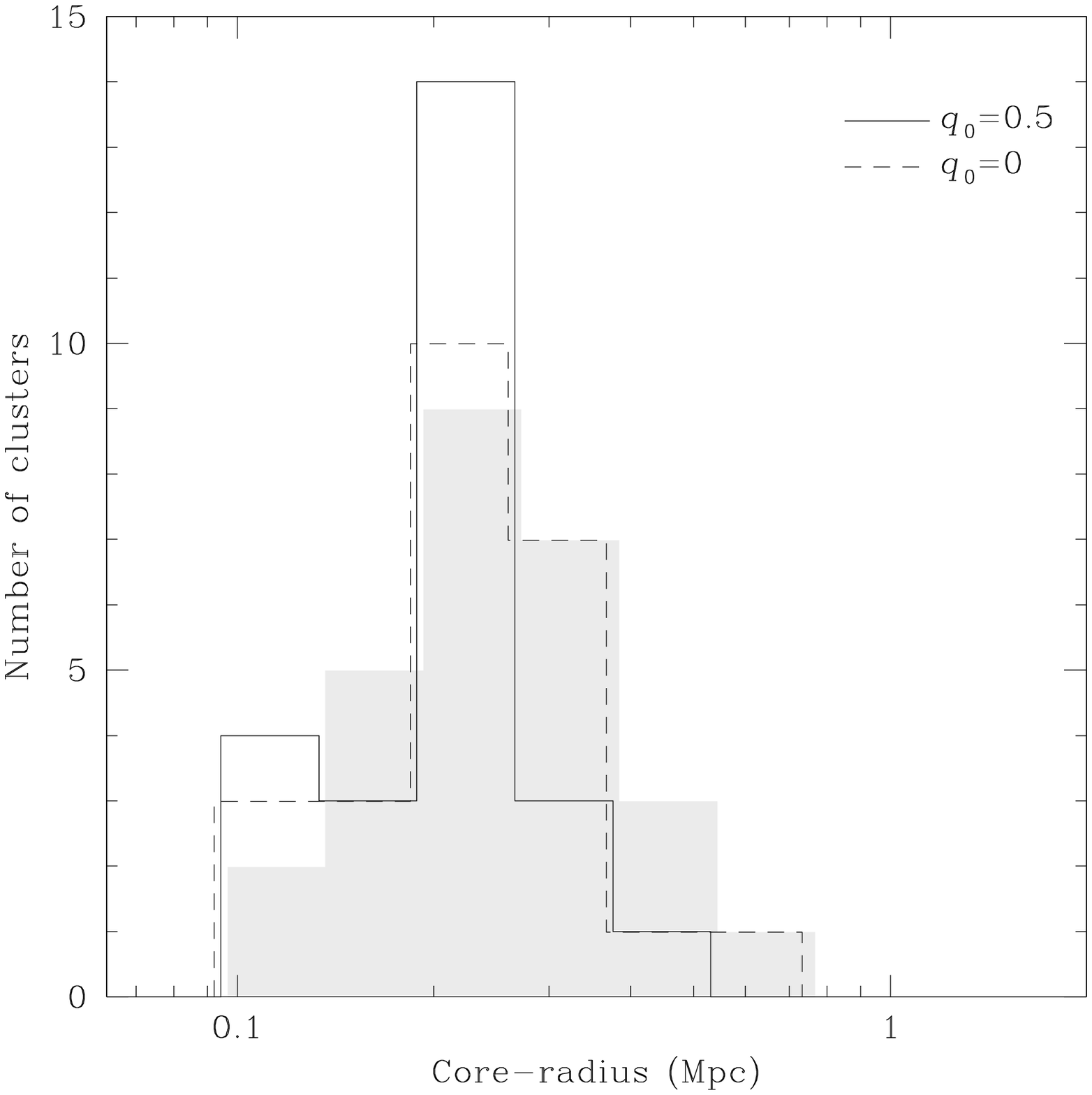}}
\vskip -15pt
\figcaption{Core-radius distribution for distant, $z>0.4$, clusters, derived
from our survey (solid and dashed histogram for $q_0=0.5$ and $q_0=0$,
respectively). The shaded histogram shows the core-radius distribution for
nearby luminous clusters from Jones \& Forman (1998). The angular resolution
limit of our survey (15\arcsec) corresponds to 120~kpc at the median
redshift of distant clusters, well below the peak of the distribution.
\label{fig:axdistr}}
\medskip

\acknowledgements

We thank M.~Markevitch for useful comments on the manuscript. Financial
support was provided by the Smithsonian Institution and NAS8-39073 contract.
HQ acknowledges support from FONDECYT grant 8970009 and the award of
Presidential Chair in Science.

\end{document}